\documentclass[aps,prb,preprint,superscriptaddress]{revtex4-2}

\usepackage{amsmath}
\usepackage{amssymb}
\usepackage{mathtools}
\usepackage[
locale=DE,
separate-uncertainty=true,
per-mode=symbol-or-fraction,
output-decimal-marker = {.},
]{siunitx}

\usepackage{lineno}
\begin{document}

%\preprint{}

\title{Quantum Resource Theory of Lasers}


\author{Yannik Brune}
\affiliation{Department of Physics, Technische Universität Dortmund, 44227 Dortmund, Germany}
\author{Marius Cizauskas}
\affiliation{Department of Physics, Technische Universität Dortmund, 44227 Dortmund, Germany}
\affiliation{School of Mathematical and Physical Sciences, University of Sheffield, Sheffield, UK}
\author{Marc Aßmann}
\affiliation{Department of Physics, Technische Universität Dortmund, 44227 Dortmund, Germany}
\affiliation{Dortmund Center for Advanced Exploration of Dynamics Across Limits Using
Spectroscopy (DAEDALUS), Technische Universität Dortmund, 44227 Dortmund, Germany}
%\thanks{}


\date{\today}

\begin{abstract}
Lasers serve as the fundamental workhorses of photonic quantum technologies, with perfectly coherent light fields being essential for many protocols that generate nonclassical light, implement coherent control schemes, and initialize qubits.
However, no laser is absolutely ideal and the implications of deviations from perfect coherence in quantum technological tasks remain unclear.
In this study, we theoretically and experimentally explore the quantum coherence properties of lasers from a resource theory perspective, establishing a significant connection between photonics, quantum optics, and quantum information science. We demonstrate that the maximum achievable quantum coherence for laser light is constrained by spontaneous emission and the purity of the dephased laser field state.
As a critical example application in quantum information protocols, we show that the quantum coherence of a laser field with a given mean photon number directly governs the maximum purity attainable when initializing a qubit in a superposition state through resonant driving.
Our findings are highly relevant for bridging applied physics and engineering with integrated photonic quantum technologies and resource theories, paving the way for reliable benchmarking of various coherent light sources for applications in photonics and quantum protocols.
\end{abstract}

%\keywords{}

\maketitle

\section{Introduction}
Combined advancements in theoretical physics, experimental physics, and engineering have historically driven human progress, as illustrated by the connection between the industrial revolution, thermodynamics, and the steam engine.
There is growing interest in the potential for quantum technologies to similarly transform society, with quantum resource theories potentially serving a role analogous to that of thermodynamics during industrialization.
Quantum resource theories \cite{Chitambar2019} establish a clear hierarchy that ranks quantum states based on their usefulness for various concrete tasks, including quantum computation and state discrimination.
This framework offers a robust quantitative approach for analyzing processes within quantum technologies.
Typically, states that are challenging to produce are deemed resourceful, whereas easily accessible states are not. 
From a theoretical standpoint, nonclassical states of light fields are often classified as resourceful, while coherent states are not; however, this classification may not hold from an applied physics perspective.

Since the first observation of lasing in 1960 \cite{Maiman1960}, numerous laser designs have been developed and adapted for a wide range of applications.
Nevertheless optimizing the coherence of novel laser designs remains a timely and relevant challenge \cite{Santis2014,Kim2016}.
While it is desirable to achieve high intensity, high coherence, low cost, and compact laser designs, realizing all these features simultaneously is often impractical.
Consequently, careful trade-offs among these properties are necessary to create the ideal laser for a specific application.
One significant challenge in benchmarking lasers is determining the acceptable level of deviation from perfect coherence for a given task.
This can be framed within a resource theory context, where highly coherent states are considered resourceful while states lacking coherence are not. 
In this work, we will develop a resource theory tailored for lasers used in quantum technologies. 
The most appropriate framework for this purpose is the resource theory of quantum coherence, which examines the operational values of quantum states exhibiting superpositions in a designated basis compared to incoherent states devoid of such superpositions \cite{Baumgratz2014,Marvian2016}.
The resourcefulness of states exhibiting superpositions is intuitively evident in systems defined on finite-dimensional Hilbert spaces, such as qubits, where superpositions are generally regarded as inherently non-classical. 
This intuition is confirmed by various resource theories demonstrating that, for these systems, the purity of a state constrains the amount of quantum coherence that can be generated through unitary operations \cite{Streltsov2018} and that quantum coherence may be converted into quantum discord and entanglement \cite{Ma2016,Zhu2017}.
In contrast, the resourcefulness of superpositions in systems defined on infinite-dimensional Hilbert spaces, such as light fields, is less intuitive. 
Some common misunderstandings arising most commonly in optics and semiconductor physics often stem from historical definitions of non-classicality that emphasize sub-Poissonian photon number statistics \cite{Kimble1977,Diedrich1987}, entanglement \cite{Horodecki2009}, or squeezing \cite{Walls1983} as indicators of optical non-classicality. 
While these measures indeed correlate with useful resources in quantum technologies — such as photonic Fock states forming the foundation for linear optical quantum computing \cite{Knill2001,Wang2019,Maring2024}, entanglement serving as a crucial resource for device-independent quantum key distribution \cite{Acin2006,Vazirani2014,Zapatero2023}, and squeezing being valuable for teleportation and entanglement generation \cite{Braunstein1998,Furusawa1998,Scheel2001}, Gaussian Boson sampling \cite{Hamilton2017,Zhong2020}, and interferometry \cite{Schnabel2017} — the broader implications of coherence in infinite-dimensional systems warrant further exploration.
In the early development of resource theories, the wide range of applications for states that meet optical non-classicality criteria occasionally led to the misconception that optical non-classicality is equivalent to resourcefulness and non-classicality in information science. 
This assumption can be easily refuted; for instance, optically classical vacuum states serve as effective resources for semi-device-independent quantum random number generation \cite{Avesani2018,Bruynsteen2023}, and even incoherent thermal light can be employed for the same purpose \cite{Qi2017,Thewes2019}. 
Moreover, it has been demonstrated that quantum computational speed-ups can be achieved using states that are entirely classical in optics \cite{Agudelo2013,Shahandeh2017}. 
Furthermore, it has been conclusively shown that the definitions of non-classicality in optics and quantum information are not equivalent \cite{Ferraro2012,Lund2023}. 
There is now clear evidence that optically classical light field states can be resourceful and play a significant role in quantum technologies. 
For tasks such as coherent control of qubits, optically classical fields have already become the standard choice \cite{Cirac1995,Turchette1998,Ahn2023}. 
Despite this recognition, the unwarranted conflation of non-classicality in optics with its counterpart in quantum information has resulted in a relative scarcity of studies focusing on the resourcefulness of optically classical states.

In this work, we address part of this gap by investigating the quantum coherence of displaced thermal states — general optically classical states characterized by a non-negative Wigner function that do not show any squeezing. 
These states correspond to those produced in real lasers or masers, which exhibit deviations from perfect coherence due to thermal admixtures. 
Given that coherent control is a prominent task in quantum technologies and electromagnetic fields are ubiquitous in such applications, we will explore the resourcefulness of displaced thermal states through the example of coherent control of qubits.
While our focus is on electromagnetic fields in the optical domain, our discussion is broadly applicable across the entire electromagnetic spectrum.
We examine a fundamental scenario: initializing a qubit by driving it from the ground state to an equally weighted superposition between the ground and excited states, so that further manipulation of that state may be performed.
A typical example for such a system is given by electron spins in quantum dots \cite{Press2008}.
It is widely acknowledged that this process will likely be performed using optically classical electromagnetic fields for scalable real-world quantum technological applications.
Although Fock states or squeezed light could theoretically also serve this purpose, coherent states are easier to produce, more robust, and readily available.
Therefore, it seems reasonable to employ optically non-classical states rather in specialized tasks that fully utilize their properties \cite{Tiarks2016} and make use of coherent states for more common tasks.
This raises an important question regarding the necessity of a resource theory for readily accessible states.
Nonetheless, it is evident that the laser industry encompasses a vast array of laser types, with ongoing significant research aimed at optimizing their coherence properties.
Still, almost all available lasers exhibit some degree of deviation from perfect coherence. 
When considering practical laser sources for integration into large-scale quantum technologies, factors such as miniaturization, low power consumption, and operational viability under challenging environmental conditions become critical.
Thus, it is essential to determine how much deviation from ideal coherence remains acceptable before the performance of, e.g., quantum computers begins to deteriorate due to limitations imposed by non-ideal laser properties.
While coherent control has been previously examined within resource theories from a general perspective \cite{Matera2016} and in terms of optimizing pulse sequences \cite{Uhrig2007,Koch2022}, our focus here is explicitly on the sources of these control fields. 
To this end, we study a fundamental coherent control task and provide a performance quantifier that enables users to assess the expected performance of specific lasers in quantum technologies while offering manufacturers clear benchmarks for optimization.

\section{Quantum Resource Theoretical Treatment of Quantum Coherence for Classical Light sources}

In this section, we present a quantum resource theoretical framework designed to quantify the usefulness of laser light field states for protocols in quantum technologies.
Our aim is to make this discussion accessible to scientists in materials science and laser manufacturing; therefore, we will minimize theoretical details and direct readers to more comprehensive accounts on the resource theory of quantum coherence \cite{Baumgratz2014,Marvian2016,Streltsov2017,Chiribella2017,Bischof2019}, particularly regarding continuous variable resource theories \cite{Gianfelici2021,Regula2021} and the role of quantum coherence in various applications within quantum technologies \cite{Wenniger2024,Loredo2019}.

Resource theories generally focus on quantifiers that define the resourcefulness of a particular state and on free incoherent operations that cannot increase a state's resourcefulness.
In our investigation, we concentrate on quantum coherence and examine how it influences the performance of light fields in qubit initialization and general coherent optical control of qubit states.
For a state characterized by its density matrix  $\hat{\rho}$, we define its quantum coherence $\mathcal{C}(\hat{\rho})$ as follows:
\begin{equation}
    \label{eq:QCoherence}
    \begin{aligned}
        \mathcal C(\hat{\rho})=&\sum_{m,n\in\mathbb N:m\neq n}|\hat{\rho}_{m,n}|^2
        \\
        =&\|\mathrm{\hat{\rho}-\hat{\rho}_\mathrm{inc}}\|_\mathrm{HS}^2
        =\mathrm{tr}\left(\hat{\rho}^2\right)-\mathrm{tr}\left(\hat{\rho}_\mathrm{inc}^2\right),
    \end{aligned}
\end{equation}
This definition can be interpreted as the sum of the squared moduli of all off-diagonal density matrix elements in the Fock basis or equivalently as the distance between the density matrix of a state $\hat{\rho}$ and its closest incoherent counterpart $\hat{\rho}_\mathrm{inc}$, using the squared Hilbert-Schmidt norm as the distance measure.
It should be noted that when employing a broad definition of incoherent operations, one may construct light fields with density matrices for which the chosen distance measure can yield misleading results, compromising the monotonicity of quantum coherence as a resource quantifier \cite{Rana2016}.
To avoid this issue, we will limit our focus to states that can be expressed as displaced thermal states and will consider a restricted set of genuinely incoherent free operations, such as transmission through vacuum or simple media and coupling to a qubit in its ground state.
For this selection of states and incoherent operations, our distance measure remains a reasonable metric \cite{deVicente2017,Lueders2021}.
For any displaced thermal state, constructing the closest incoherent counterpart to a state $\hat{\rho}$ is straightforward.
Incoherent states are characterized by density matrices that are diagonal in the chosen basis.
The closest incoherent state to $\hat{\rho}$ is obtained by removing all off-diagonal elements from its density matrix. 
This incoherent state can be thought of as the phase-randomized version of the initial state, where all phase information has been effectively erased.

Our choice of the squared Hilbert-Schmidt norm as a quantifier for quantum coherence establishes a direct link to the purity $\mathcal P(\hat{\rho})$ of our state of interest:
\begin{equation}
    \label{eq:Purity1}
        \mathcal P(\hat{\rho})=\mathrm{tr}\left(\hat{\rho}^2\right).
\end{equation}
Consequently, the quantum coherence of a state corresponds directly to the difference between the purities of the initial state and its incoherent counterpart:
\begin{equation}
    \label{eq:Purity2}
        \mathcal C(\hat{\rho})=\mathcal P(\hat{\rho})-\mathcal P(\hat{\rho}_\mathrm{inc}).
\end{equation}
This relationship implies that quantum coherence is diminished during dephasing, which aligns with intuitive expectations.
For our purposes, it will be beneficial to express the equation in terms of linear entropy or mixedness $\mathcal M(\hat{\rho})=1-\mathcal P(\hat{\rho})$ of $\hat{\rho}$:
\begin{equation}
    \label{eq:Purity3}
        \mathcal C(\hat{\rho})=1-\mathcal M(\hat{\rho})-\mathcal P(\hat{\rho}_\mathrm{inc})
\end{equation}
or equivalently,
\begin{equation}
    \label{eq:CoherenceFinal}
        \mathcal C(\hat{\rho})+\mathcal M(\hat{\rho})+\mathcal P(\hat{\rho}_\mathrm{inc})=1
\end{equation}

This equation serves as the central theoretical foundation from which we will discuss the usefulness of lasers in coherent control schemes.
First, it clearly illustrates how obtainable quantum coherence is constrained by both the mixedness of a state and the purity of its corresponding incoherent dephased state.
Second, it enables us to construct a simple toy model for the onset of lasing, providing an intuitive understanding of how $\mathcal C(\hat{\rho})$, $\mathcal M(\hat{\rho})$ and $\mathcal P(\hat{\rho}_\mathrm{inc})$ contribute to the lasing process.

\section{A Resource-theoretical Laser Model}\label{sec:LaserModel}
In this section, we examine the formation of a single lasing mode from the perspective of resource theory, specifically focusing on a displaced thermal state.
The discussion is designed to be pedagogical, dividing the formation process into four distinct steps that emphasize the effects of the three quantities outlined in equation (\ref{eq:CoherenceFinal}).
While it is important to note that these processes do not occur independently in a real laser and cannot be completely separated, we will demonstrate how to isolate individual contributions by analyzing the density matrix or phase space representation corresponding to the lasing mode. 
Additionally, we will illustrate what the Wigner function looks like for each of the intermediate states.

\begin{figure}
    \centering
    \includegraphics[width=\textwidth]{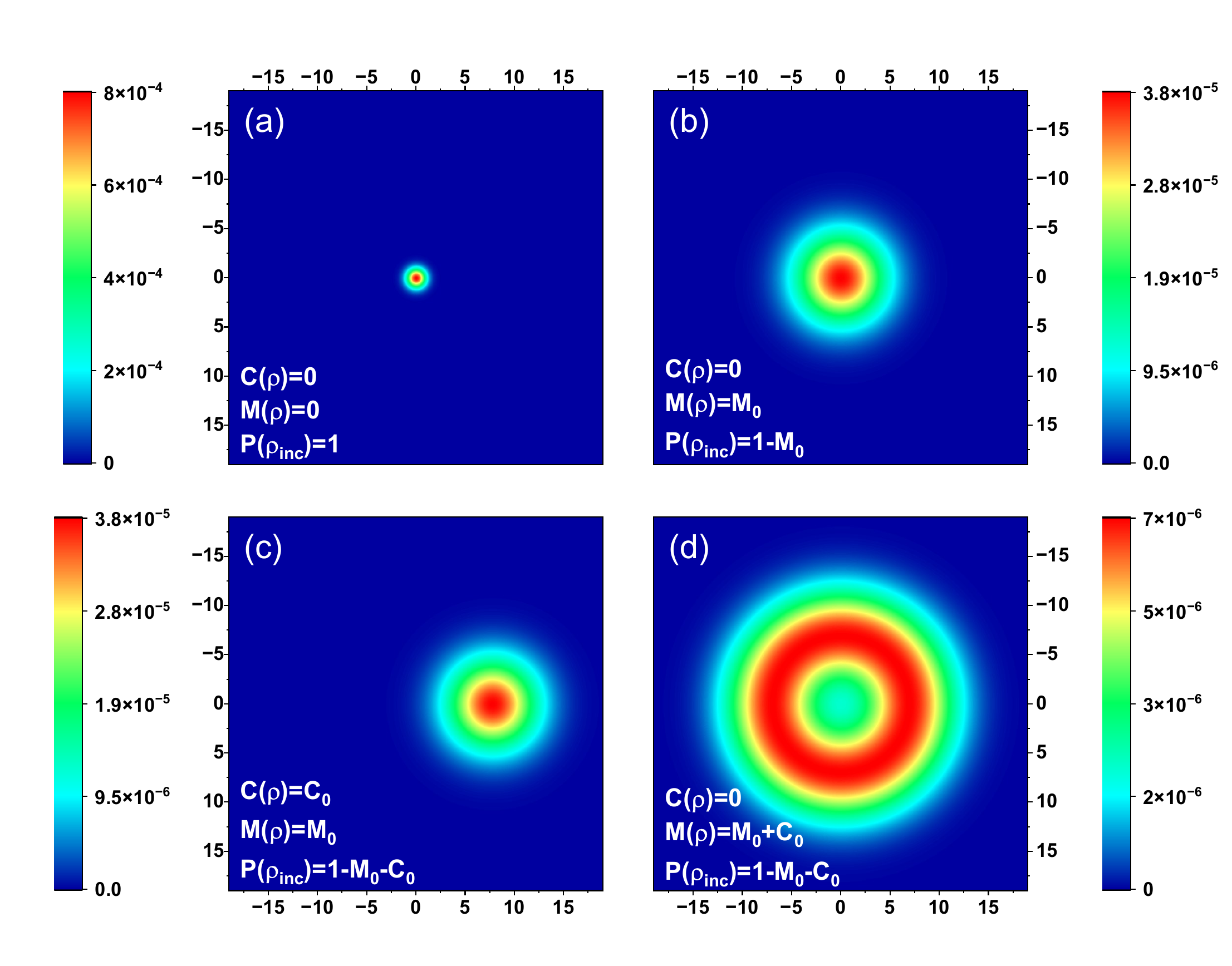}
    \caption{Wigner functions for the intermediate states arising in the resource-theoretical laser model. (a) vacuum state. (b) Thermal state with $\langle N\rangle_\mathrm{th}$=10. (c) Displaced thermal state with $|\alpha_0|=\sqrt{60}$. (d) Dephased displaced thermal state.}
    \label{fig:Wigner_Model}
\end{figure}

We begin with the vacuum state $\hat{\rho}_\mathrm{vac}$, whose Wigner function is depicted in Fig.\ref{fig:Wigner_Model}(a). 
In the Fock basis, this state represents a pure state with exactly zero photons.
As it is a pure state, its mixedness vanishes: $\mathcal M(\hat{\rho}_\mathrm{vac}) = 0$.
Furthermore, its density matrix does not exhibit any off-diagonal elements, resulting in $\mathcal C(\hat{\rho}_\mathrm{vac}) = 0$. 
Since $\hat{\rho}_\mathrm{vac}$ contains no off-diagonal elements, its closest incoherent counterpart remains $\hat{\rho}_\mathrm{vac}$ itself.
Consequently, as previously discussed, this pure state leads to $\hat{\rho}_\mathrm{vac,inc}=\hat{\rho}_\mathrm{vac}$ and $\mathcal P(\hat{\rho}_\mathrm{inc})=1$.

In the next stage of our model, the laser mode becomes populated through spontaneous emission, allowing stimulated emission to subsequently occur. 
We model this initial incoherent spontaneous population by introducing a thermal component, resulting in a thermal state density matrix $\hat{\rho}_\mathrm{th}$.
An exemplary Wigner function for such a state is shown in Fig.\ref{fig:Wigner_Model}(b). Thermal states are incoherent states with density matrices that lack off-diagonal elements. 
Consequently, the light field is in a mixed state composed of pure Fock state density matrices $|i\rangle\langle i|$ weighted by their respective probabilities $p_i$:
\begin{equation}
    \label{eq:FockDM}
        \hat{\rho}_\mathrm{th}=\sum_i p_i |i\rangle\langle i|.
\end{equation}
For a given mean thermal photon number $\langle N\rangle_\mathrm{th}$, these probabilities follow a Bose-Einstein distribution:
\begin{equation}
    \label{eq:BEDist}
        p_n=\frac{\langle N\rangle_\mathrm{th}^n}{(\langle N \rangle_\mathrm{th} +1)^{n+1}}.
\end{equation}

As this density matrix still lacks off-diagonal elements, it contains no quantum coherence: $\mathcal C(\hat{\rho}_\mathrm{th})=0$.
The purity and mixedness of a thermal state are well established to depend solely on the mean thermal photon number \cite{Lueders2021}. 
The mixedness of the thermal state is given by:
\begin{equation}
    \label{eq:thMixedness}
       \mathcal M(\hat{\rho}_\mathrm{th})=\frac{2 \langle N \rangle_\mathrm{th}}{2\langle N \rangle_\mathrm{th}+1}.
\end{equation}
Since the thermal state is fully incoherent, we have $\hat{\rho}_\mathrm{th,inc}=\hat{\rho}_\mathrm{th}$. 
The purity of the incoherent state then follows as:
\begin{equation}
    \label{eq:thIncPurity}
\mathcal P(\hat{\rho}_\mathrm{th,inc})=1- \mathcal M(\hat{\rho}_\mathrm{th})=\frac{1}{2\langle N \rangle_\mathrm{th}+1}.
\end{equation}

In the next step of our laser model, we incorporate the development of the coherent component of the light field through stimulated emission. 
This is achieved by applying a coherent displacement operator $\hat{D}(\alpha_0)$ to the thermal state, where $\alpha_0$  represents the coherent displacement in phase space. 
Consequently, we obtain the density matrix of a displaced thermal state given by $\hat{\rho}_\mathrm{DT}=\hat{D}(\alpha_0)\hat{\rho}_\mathrm{th}\hat{D}^\dagger(\alpha_0)$. 
The corresponding Wigner function is illustrated in Fig.\ref{fig:Wigner_Model}(c).
Coherent displacement is a unitary operation that does not affect the mixedness of the state; thus, we have $\mathcal M(\hat{\rho}_\mathrm{DT})= \mathcal M(\hat{\rho}_\mathrm{th})=\frac{2 \langle N \rangle_\mathrm{th}}{2\langle N \rangle_\mathrm{th}+1}$. 
The quantum coherence $ \mathcal C(\hat{\rho}_\mathrm{DT})$ will acquire a finite value, which is our primary focus to determine. 
An important observation is that the maximum quantum coherence achievable is constrained by the level of mixedness introduced in the second stage of the laser model.
This presents a significant limitation; for instance, even a seemingly small thermal contribution of $\langle N \rangle_\mathrm{th}=0.5$ photons reduces the maximal obtainable quantum coherence from 1 to 0.5, regardless of how large the coherent displacement — and consequently, the coherent photon number — may be.
It may seem surprising that such small thermal components can have such substantial effects.
However, this finding aligns with earlier experimental data indicating that the quantum coherence of polariton condensates diminishes significantly above the condensation threshold \cite{Lueders2021}.
Since the displaced thermal state now possesses finite quantum coherence, removing all off-diagonal density matrix entries results in a closest incoherent state $\hat{\rho}_\mathrm{DT,inc}$ that actually differs from the original state.
The purity $\mathcal P(\hat{\rho}_\mathrm{DT,inc})$ of this incoherent state can be calculated straightforwardly \cite{Lueders2021}:
\begin{equation}
    \label{eq:DTIncPurity}
\mathcal P(\hat{\rho}_\mathrm{DT,inc})=\frac{\exp[- 2|\alpha_0|^2 /(2 \langle N \rangle_\mathrm{th}+1 )]}{2 \langle N \rangle_\mathrm{th}+1} I_0\left[\frac{2|\alpha_0|^2}{2 \langle N \rangle_\mathrm{th}+1 }\right],
\end{equation}
where $I_0(x)=\frac{1}{2\pi}\int_0^{2\pi} d\varphi\,e^{x\,cos(\varphi)}$ represents the zeroth modified Bessel function of the first kind.
As the final step in our laser model, we consider the effects of dephasing on the prepared state.

To simulate this process, we remove all off-diagonal matrix elements from $\hat{\rho}_\mathrm{DT}$, resulting in the state $\hat{\rho}_\mathrm{deph}=\hat{\rho}_\mathrm{DT,inc}$, which is simply the closest incoherent state to our displaced thermal state of interest.

The Wigner function for this state is depicted in Fig.\ref{fig:Wigner_Model}(d).
Since this state is already incoherent, it serves as its own closest incoherent counterpart, allowing us to easily determine $\mathcal P(\hat{\rho}_\mathrm{deph,inc})$:
\begin{equation}
    \label{eq:DephIncPurity}
\mathcal P(\hat{\rho}_\mathrm{deph,inc})= \mathcal P(\hat{\rho}_\mathrm{DT,inc})
\end{equation}
Due to the absence of off-diagonal matrix elements, we also have $\mathcal C(\hat{\rho}_\mathrm{deph})=0$. 
This indicates that during the dephasing process, the quantum coherence of the displaced thermal state has been converted into additional mixedness, aligning with intuitive expectations:
\begin{equation}
    \label{eq:DephMixedness}
\mathcal M(\hat{\rho}_\mathrm{deph})= \mathcal M(\hat{\rho}_\mathrm{DT})+ \mathcal C(\hat{\rho}_\mathrm{DT}).
\end{equation}
It is important to note that both $\mathcal P(\hat{\rho}_\mathrm{DT,inc})$ and $ \mathcal M(\hat{\rho}_\mathrm{DT})$ are fully determined by their thermal and coherent photon number contributions, $\langle N \rangle_\mathrm{th}$ and $|\alpha_0|^2$, respectively. 
Interestingly, this implies that identifying these contributions — easily obtained from experimental measurements of the dephased state — is sufficient for evaluating the quantum coherence of the displaced thermal state.
At first glance, this conclusion may seem nontrivial as it assumes no partial dephasing occurs in the initial displaced thermal state and that all mixedness arises solely from the thermal component. 
However, this assumption holds true in most relevant application scenarios but necessitates a careful discussion regarding the meaning of quantum coherence within coherent control contexts.

In later sections of this manuscript, we will explicitly address how this assumption is satisfied across a wide range of scenarios we call "interferometric", which involve light beams derived from a single light field.
We will also discuss situations where this assumption may break down when synchronization of different light fields is an issue.
Next, we will demonstrate experimentally the intricate relationships among quantum coherence, mixedness, and purity in dephased states before exploring how quantum coherence influences performance in coherent control applications.

\section{Experiment}
\begin{figure}
    \centering
    \includegraphics[width=\textwidth]{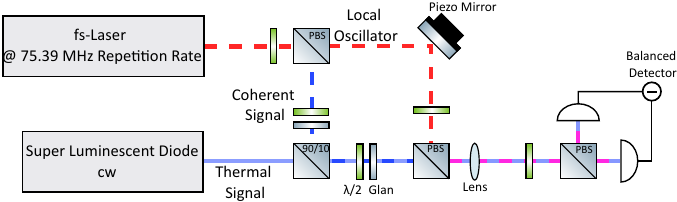}
    \caption{Experimental scheme for generating and analyzing displaced thermal states of tailored composition. A fs-laser acts as source of both local oscillator (LO) and coherent signal, while a super-luminescent diode (SLD) provides the thermal signal \cite{Doronin2019}. The displaced thermal signal state is created by mixing both input signals on a beamsplitter. Its field quadratures are measured using a 4-port homodyne detector. The quadrature sampling axis is driven by a piezo mirror.}
    \label{fig:experimental_setup}
\end{figure}
In the previous section, we introduced a quantum resource theoretical toy model that describes the fields emitted from lasers as displaced thermal states.
In the next step, we aim to validate this model experimentally.
To achieve this, we prepare displaced thermal states with varying thermal and coherent contributions and reconstruct their density matrices in the Fock basis using homodyne detection and quantum state tomography. 
We then extract the central quantities utilized in our model directly from these density matrices and compare the results with predictions derived from the displaced thermal state model.
To obtain the density matrix of the displaced thermal states experimentally, we employed a homodyne detection setup, schematically depicted in Fig. \ref{fig:experimental_setup}.
 
 A femtosecond-pulsed laser operating at $\SI{830}{\nano\metre}$ served both as the source for the coherent part of the artificial signal and as the local oscillator, enabling a quadrature sampling rate of $\SI{75.39}{\mega\hertz}$ and an intrinsic time resolution of $\SI{1}{\pico\second}$. 
 Since both the local oscillator and coherent signal originated from the same laser source, there was a stable phase relationship between them.
   
We created a displaced thermal state by overlapping this coherent signal with the thermal emission from a super-luminescent diode, also at $\SI{830}{\nano\metre}$.
The combined state was measured by overlapping it with the local oscillator before being sent to a balanced detector. 
The relative phase between the signal and the local oscillator was scanned using a piezo mirror.   
    
Subsequently, we reconstructed the density matrix of the light field using a maximum-likelihood approach as introduced by Lvovsky et al. \cite{Lvosky_2004}. 
The required phase information was obtained by monitoring the mean signal quadrature at the piezo sweep frequency. 
Finally, we constructed the Wigner function of the state from its density matrix using pattern functions \cite{Leonhardt1995}.
   
To ensure precise control over both thermal and coherent contributions in our displaced thermal state during experimentation, we implemented a fast in-situ version of our algorithms.
This approach allowed us to monitor both the density matrix of the light field and its Wigner function in real-time throughout our measurements, enabling reliable adjustments to both components while maintaining a constant mean photon number.
  
To extract the central physical quantities of our resource-theoretical model from the experimental data, we utilized both the reconstructed density matrices and the phase space distributions of the light fields.
The quantities  $\mathcal C(\hat{\rho})$,  $ \mathcal M(\hat{\rho})$ and $ \mathcal P(\hat{\rho}_\mathrm{inc})$ were directly obtained from the density matrices.
Meanwhile, the thermal and coherent photon numbers $\langle N \rangle_\mathrm{th}$ and $|\alpha_0|^2$ were determined from the width and displacement of the phase space distributions, respectively.
 \begin{figure}
    \centering
    \includegraphics[width=\textwidth]{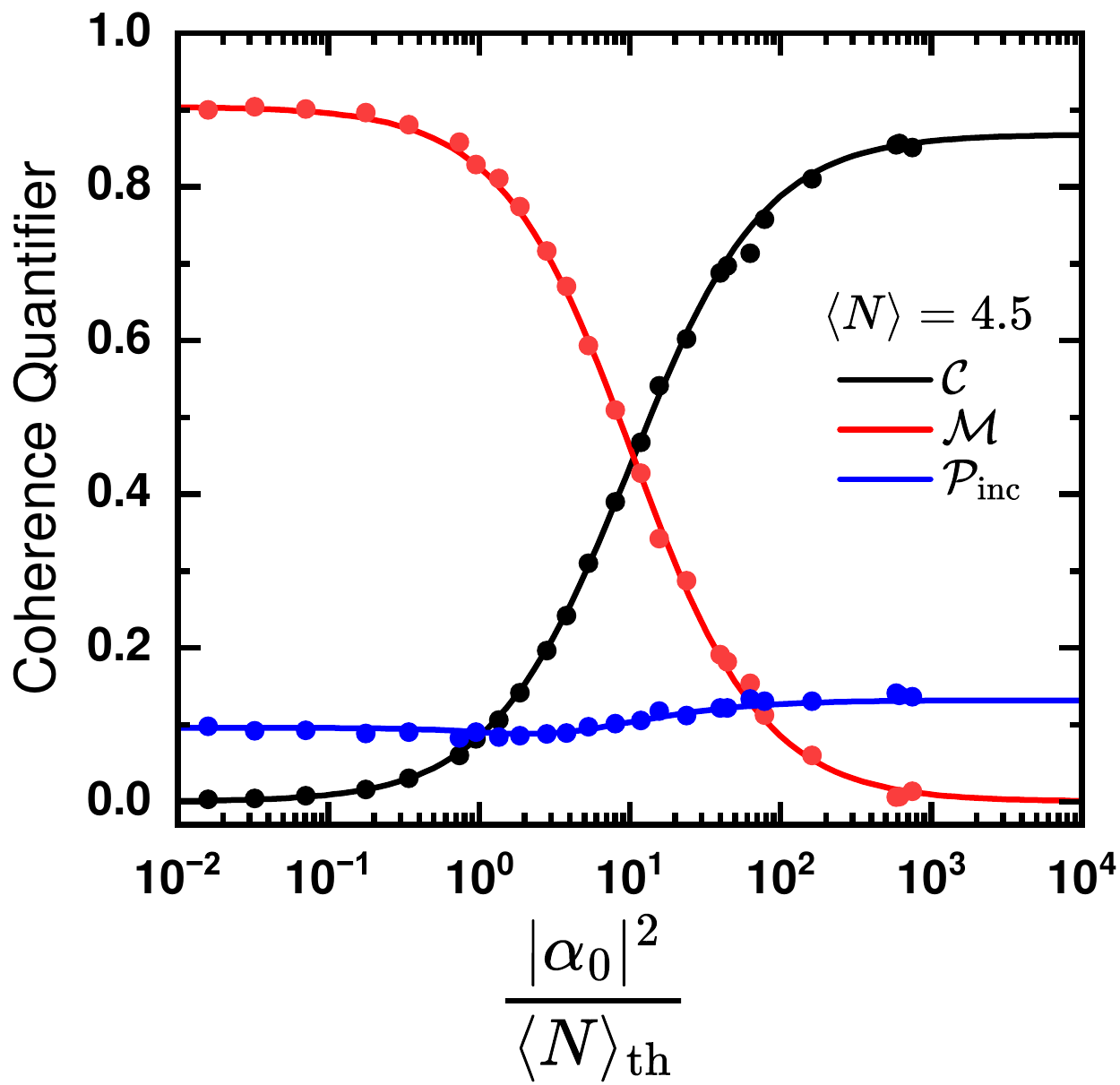}
    \caption{Quantum coherence (black), mixedness (red) and purity of the closest incoherent state (blue) for displaced thermal states with $\left<N\right> = 4.5$ with varying ratio between coherent and thermal photon numbers. Dots denote experimental results, while solid lines represent the theoretical prediction.} 
    \label{fig:CMPinc_ratio}
\end{figure}

The results for a mean photon number of 4.5 are presented in Fig. \ref{fig:CMPinc_ratio}.
The lines represent the theoretical predictions, obtained by starting with a thermal density matrix $\hat\rho_\mathrm{th}$ corresponding to a thermal photon number of $\langle N\rangle_\mathrm{th}$. 
The displaced thermal state was then constructed by applying the displacement operator $\hat{D}(\alpha_0)$, resulting in a coherent contribution of $|\alpha_0|^2$ to the photon number: $\hat{\rho}_\mathrm{DT}=\hat D(\alpha_0)\hat{\rho}_\mathrm{th} \hat{D^\dagger}(\alpha_0)$. 
The quantities $\mathcal{C}$ and $\mathcal{M}$ were directly evaluated from this state.
We then subsequently removed all off-diagonal matrix elements to assess $\mathcal P_\mathrm{inc}$.
We find excellent agreement between theory and experiment, highlighting both the reliability of our model and the precision with which we can control our setup during measurements using in-situ quantum state tomography.

\begin{figure}
    \centering
    \includegraphics[width=\textwidth]{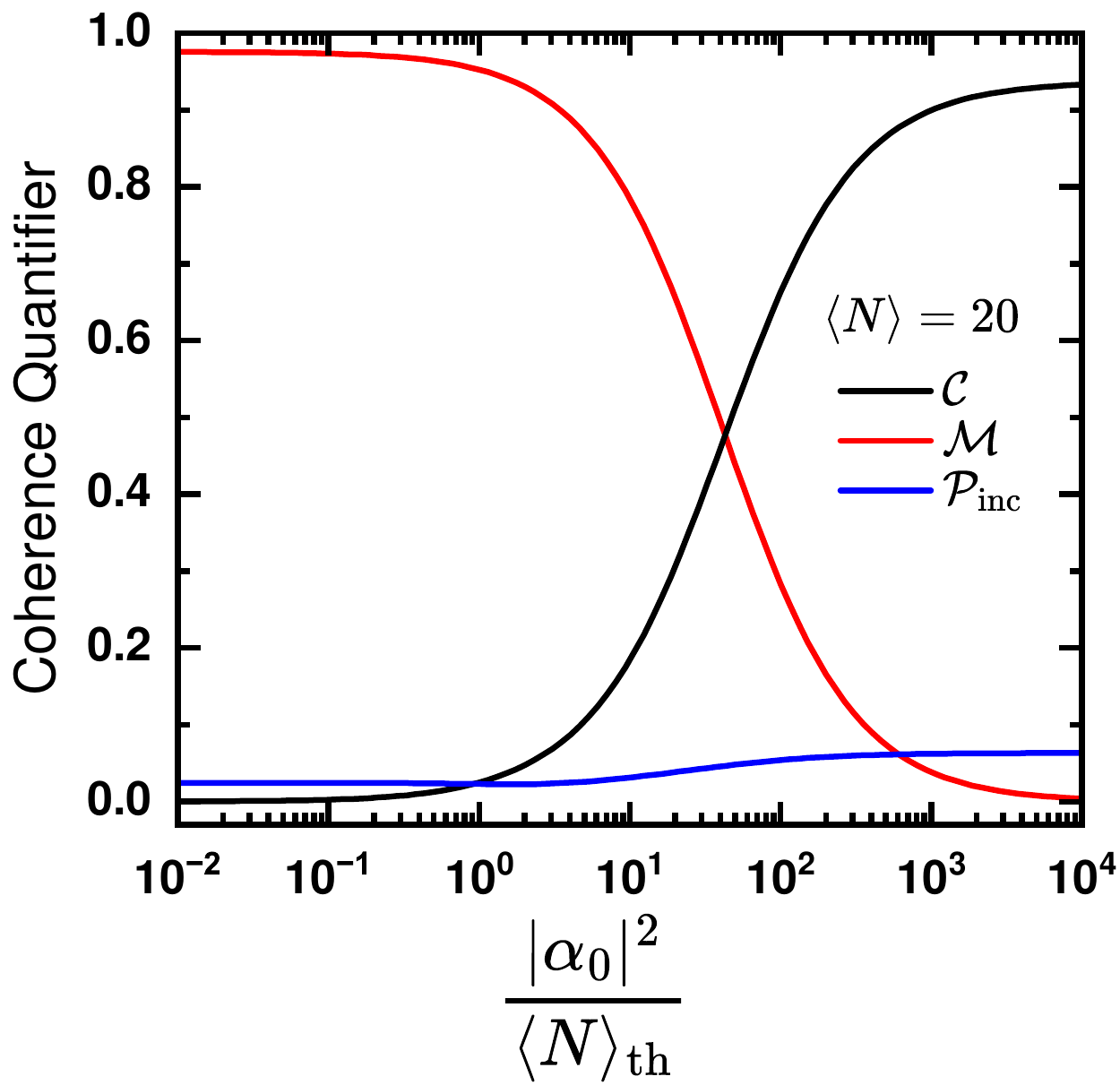}
    \caption{Quantum coherence (black), mixedness (red) and purity of the closest incoherent state (blue) for displaced thermal states with $\left<N\right> = 20$ with varying ratio between coherent and thermal photon numbers.} 
    \label{fig:CMPinc_ratioN20}
\end{figure}

There is a clear transition from predominant mixedness to predominant quantum coherence as the relative contribution of the coherent field increases.
At high coherent fractions, $\mathcal C$ approaches a constant value slightly below unity. Although the thermal component becomes negligible in this range, it is now $\mathcal P_\mathrm{inc}$ that limits the maximum obtainable quantum coherence. 
Intuitively, one might expect that this upper bound could be increased by utilizing larger total photon numbers. Indeed, $ \mathcal P_\mathrm{inc}$ can be reduced by increasing the mean photon number, as illustrated in Fig.\ref{fig:CMPinc_ratioN20} for a mean photon number of 20.  

However, it is important to note that $ \mathcal M$ depends solely on the absolute thermal photon number. 
Consequently, the required fraction of coherent photons needed to achieve saturation levels of $\mathcal C$ shifts to higher values. 
This observation highlights a key finding of our work: the absolute thermal photon number constrains obtainable quantum coherence and should be minimized when optimizing laser designs for enhanced coherence. 
Even a small thermal contamination on the order of half a photon leads to significantly reduced values of $\mathcal C$. 
With this basic understanding of what governs the quantum coherence properties of a displaced thermal light field, we now investigate, how the quantum coherence of a light field correlates with its performance in tasks in quantum technological tasks, such as coherent control.  
  
\section{Quantum Coherence and Coherent Control}

The central element of quantum coherence lies in the off-diagonal elements of a density matrix. 
In this section, we will investigate the role that quantum coherence plays in a standard task within quantum technologies: driving a qubit from its ground state $|g\rangle$ to an equally weighted superposition between the ground state and the excited state $|e\rangle$, enabling subsequent operations on that state.
We will begin with a general discussion on quantum coherence and its implications, introduce several application scenarios, and then explicitly explore how quantum coherence relates to the performance of a light field in initializing a qubit.
For simplicity, our discussion will focus on pulsed light fields. 
However, the principles are applicable to continuous fields as well, which we will address later.

In discussions surrounding quantum coherence, two questions frequently arise: the question of a appropriate reference frames in quantum optics and the relevance of dephasing in solid-state physics. 
Both are crucial for understanding the operational meaning of quantum coherence, so we will examine them in detail.

In quantum optics, there have been considerable discussion regarding whether any coherent state possesses off-diagonal matrix elements. 
A detailed description of the emission process often involves an entangled light-matter system that results in only a mixture of Fock states when tracing out the matter component and considering ensemble averages \cite{Molmer1997,Gea1998}. 
While these discussions are intriguing from a fundamental perspective, they hold little practical significance for operational or applied settings.
The presence of off-diagonal elements indicates coherence relative to some phase reference frame. 
From a fundamental standpoint, this is akin to defining a reference frame with an absolute phase — an approach that lacks physical meaning. 
Indeed, no practically relevant experimental realization of a coherent state will exhibit non-vanishing off-diagonal matrix elements with respect to some arbitrary absolute phase.
However, in contexts such as spectroscopy, metrology, or coherent control, more meaningful reference frames exist. 
Quantum coherence can be defined relative to these frames, which we will now discuss.

\subsection{The synchronization scenario}
A common practical example is an optical amplifier, where an external classical seed light field serves as a suitable phase reference for the emitted light field. 
Neither the seed nor the emitted light field exhibits permanent coherence with respect to any arbitrarily defined external reference frame.
However, the seed can be utilized as a reliable clock that provides a reference frame for the emitted light. 
It is physically meaningful to assess how well the output reference frame aligns with the external seed reference frame, which we will refer to as the synchronization scenario.

This scenario involves studying an ensemble of pulses that are repeatedly seeded by another light field. 
In this context, quantum coherence can be degraded by two distinct mechanisms. 
First, as discussed in our laser model, any thermal component present in the emitted light field contributes to mixedness. 
Second, variations in the relative phase between the emitted light field and the seed from shot to shot further diminish the purity of the emitted state when averaged over the ensemble of all pulses.
This scenario accommodates displaced thermal states that are partially dephased due to phase fluctuations relative to the seed. 
Consequently, knowing the coherent and thermal photon number contributions of the fully dephased incoherent state is insufficient for determining quantum coherence, which now also quantifies how well the output of the optical amplifier is phase-synchronized with the seed. 
However, knowledge of these two photon numbers still enables one to calculate the maximum achievable quantum coherence.
The exact quantum coherence can be determined experimentally by using a portion of the seed beam as a local oscillator, allowing for the reconstruction of the full density matrix of the emitted light field through quantum state tomography. 
In coherent control of a qubit, this description is particularly relevant when employing the seed laser to transition the qubit from its ground state $|g\rangle$ to a superposition state $|+\rangle=\frac{1}{\sqrt{2}}(|g\rangle +|e\rangle)$ via a $\frac{\pi}{2}$ pulse, followed by applying the output from the amplifier to perform subsequent operations on the qubit — such as another $\frac{\pi}{2}$ pulse to drive it to the excited state.
While this pulse sequence can certainly be implemented, using two different beams for these operations is not common or optimal in practice due to potential noise in the relative phase between them. 
This phase noise reflects a lack of knowledge regarding the appropriate reference frame for operations on the qubit. 
We may, e.g., rotate our reference frame such that state $|+\rangle$ becomes $|-\rangle=\frac{1}{\sqrt{2}}(|g\rangle -|e\rangle)$ in this new frame, and vice versa. 
The same $\frac{\pi}{2}$ pulse that drives a qubit in state $|+\rangle$ to its excited state will instead drive a qubit in state $|-\rangle$ back to its ground state.
Of course, this issue can be addressed by transforming both pulses into the new reference frame.
However, any phase noise signifies a lack of necessary information for doing so. Nonetheless, we will not pursue this scenario in detail here since using pulses derived from different laser sources is not an optimal strategy for coherent control.

\subsection{The interferometric scenario}
A more relevant use case for coherent control is given by a pulse sequence in which a single output pulse from the amplifier is divided into two pulses.
These pulses are used to drive the qubit first from its ground state to the superposition state and then from the superposition state $|+\rangle$  to, e.g., the excited state. 
While the phase of these pulses relative to the seed laser may still be subject to noise, this is inconsequential for the coherent control task. 
The seed is not actually utilized, meaning that any noise translates into an irrelevant absolute phase shift that affects both output pulses equally.

Thus, we can simplify our scenario by considering a strong output pulse from an unseeded laser, which we split into two partial beams to create both $\frac{\pi}{2}$ pulses for driving the qubit. 
We will refer to this as the interferometric scenario. 
In this context, two identical copies of a light field serve as both the state of interest and the reference light field. 
This phase reference effectively acts as a clock, allowing it to be considered perfectly coherent with itself \cite{Wiseman2004}, thus representing a reference state of zero phase.
This perspective has led to valid criticisms emphasizing that coherence describes relationships and questioning whether it is reasonable to define a quantity solely with respect to itself without any external reference frame \cite{Bartlett2006}. 
Nevertheless, even in the absence of an external reference frame, considering relative phases suffices for establishing a meaningful definition of coherence \cite{Kae2003}.
In fact, self-referencing experiments are routinely conducted in both industrial and academic laboratories: many interferometric experiments utilize self-referencing by creating two copies of a light field that interfere with each other. 
For example, every Michelson interferometry experiment aimed at determining the classical coherence time of a light field relies on comparing that field against itself. 
We can investigate quantum coherence within a similar operational framework.
Loosely speaking, in the interferometric scenario, we evaluate how effectively a particular light field can serve as a phase reference for other fields. 
In contrast, in the synchronization scenario, we assess how well a specific light field aligns with an established phase reference.
However, there is an additional caveat. 
In classical interferometry using single modes, we must shift the two copies of the light field relative to each other — either in space or in time — to obtain meaningful results.
Interfering two identical copies of the same light field without any delay is equivalent to performing a first-order coherence measurement at zero temporal and spatial delay, which yields a value of 1 for every single-mode light field.
In contrast, utilizing these same two identical unshifted copies as a signal and phase reference to measure the quantum coherence of the signal will produce significantly different results depending on the specific states of the light fields. 
We previously discussed this behavior: any thermal component introduces mixedness that reduces the quantum coherence of the state.
Assuming ideal experimental conditions and neglecting the noise introduced by splitting the beam, both beams will be perfectly in phase.
Thus, additional dephasing present in the synchronization scenario is absent. 
In this case, all mixedness in the displaced thermal state arises solely from its thermal component. 
Interestingly, this implies that measuring only the thermal and coherent contributions to the photon number of the dephased incoherent state is sufficient to fully determine the quantum coherence of the displaced thermal state.

From an experimental point of view, this represents a significant simplification since measuring photon numbers is considerably easier than conducting full quantum state tomography as required in the synchronization scenario. 
For instance, it suffices to measure the phase-averaged Husimi Q function of the signal light field, which can be achieved using a local oscillator without any fixed phase relationship with respect to the signal.
Consequently, we are not restricted to signals derived from a local oscillator.
We can assess the quantum coherence of any single-mode light field. 
Returning to our example of driving a qubit with two $\frac{\pi}{2}$ pulses, this means that the interferometric scenario delineates intrinsic limitations regarding how effectively a qubit state can be prepared using a given light field.

\subsection{The dephasing scenario}
One may also envision intermediate scenarios that bridge those previously discussed.
For instance, we can again examine an unseeded laser pulse split into two partial beams while relaxing the assumption of ideal experimental conditions. 
In this case, the paths taken by the two beams may differ significantly, potentially exposing them to air or mechanical fluctuations that degrade the relative phase of the light fields.
Along similar lines, we may consider driving the qubit with two or more consecutive pulses emitted from the same laser. 
As the waiting time between consecutive pulses increases, the relative phase between them becomes increasingly randomized due to dephasing processes occurring within the laser. 
In such cases, it is common to investigate how dephasing grows as a function of time or distance.
We can therefore view this scenario as a modified synchronization scenario, where quantum coherence is still considered between a signal of interest and a reference beam.
In this scenario, a time-delayed or spatially shifted copy of the signal serves as the reference. 
The dephasing of quantum coherence in this context shares some similarities with the information obtained from a conventional $g^{(1)}$ first-order coherence measurement \cite{Thewes2020,Lueders2023}. 
However, a significant difference arises: for zero temporal delay and spatial shift, this situation reverts to the interferometric scenario, yielding the intrinsic quantum coherence specific to the state. 
In contrast, a $g^{(1)}$-measurement will simply produce a fixed value of 1 for any single-mode light field.
In practice, any light field state will experience dephasing over some characteristic timescale. 
For delays much shorter than this dephasing time, one will obtain the state with maximal coherence.
Conversely, for delays much longer than the dephasing time, one would observe a fully dephased state. 
This type of information is particularly valuable in semiconductor physics for understanding dynamic processes and interactions occurring within the laser or along the path of the light field.
However, unless dephasing occurs on timescales shorter than the duration of the pulse, it becomes irrelevant for characterizing the light source itself concerning coherent control from a resource perspective. 
Dephasing still remains significant when characterizing not only light sources but instead entire devices that include optical elements and beam propagation paths. Importantly, this means that one can evaluate the performance of a given light source independently from that of any device built around it.

\section{Quantum Coherence and State Preparation}
Following the qualitative discussion of various scenarios involving quantum coherence and their typical applications in different fields in the last section, we will now introduce a quantitative measure that describes the performance of a light field in preparing a qubit in a superposition state. 
The natural figure of merit for this purpose is the purity of the qubit state after tracing out the light field degrees of freedom. 
Our focus will be on the interferometric and dephasing scenarios.

Specifically, we will first investigate the maximum purity of the qubit superposition state achievable with a Rabi pulse possessing a certain degree of quantum coherence, independent of any interactions between the pulses and their environment or other perturbations. 
We will then separately examine the influence of dephasing.
To this end, we consider the Jaynes-Cummings Hamiltonian in the rotating wave approximation:
\begin{equation}
    \label{eq:Hamiltonian}
\hat{H}=\frac{1}{2}\hbar\omega_\mathrm{q} \hat{\sigma}_z+\hbar\omega_\mathrm{c} \hat{a}^\dagger \hat{a}+g(\hat{a}^\dagger\hat{\sigma}+\hat{a}\hat{\sigma}^\dagger),
\end{equation}
where $\hat{a}^\dagger$ and $\hat{a}$ denote the creation and annihilation operators for the light field mode, respectively.
$\hat{\sigma}^\dagger$ and $\hat{\sigma}$ are the raising and lowering operators for the qubit.
$\hat{\sigma}_z$ is the qubit inversion operator and $\omega_\mathrm{q}$ and $\omega_\mathrm{c}$ correspond to the energy level difference of the qubit states and the energy of the photon mode, respectively. 
 The parameter $g$ represents the light-matter coupling strength. 
 For our analysis, we will consider resonant driving conditions, such that $\omega_\mathrm{q}=\omega_\mathrm{c}$.

\begin{figure}
        \centering
        \includegraphics[width=\textwidth]{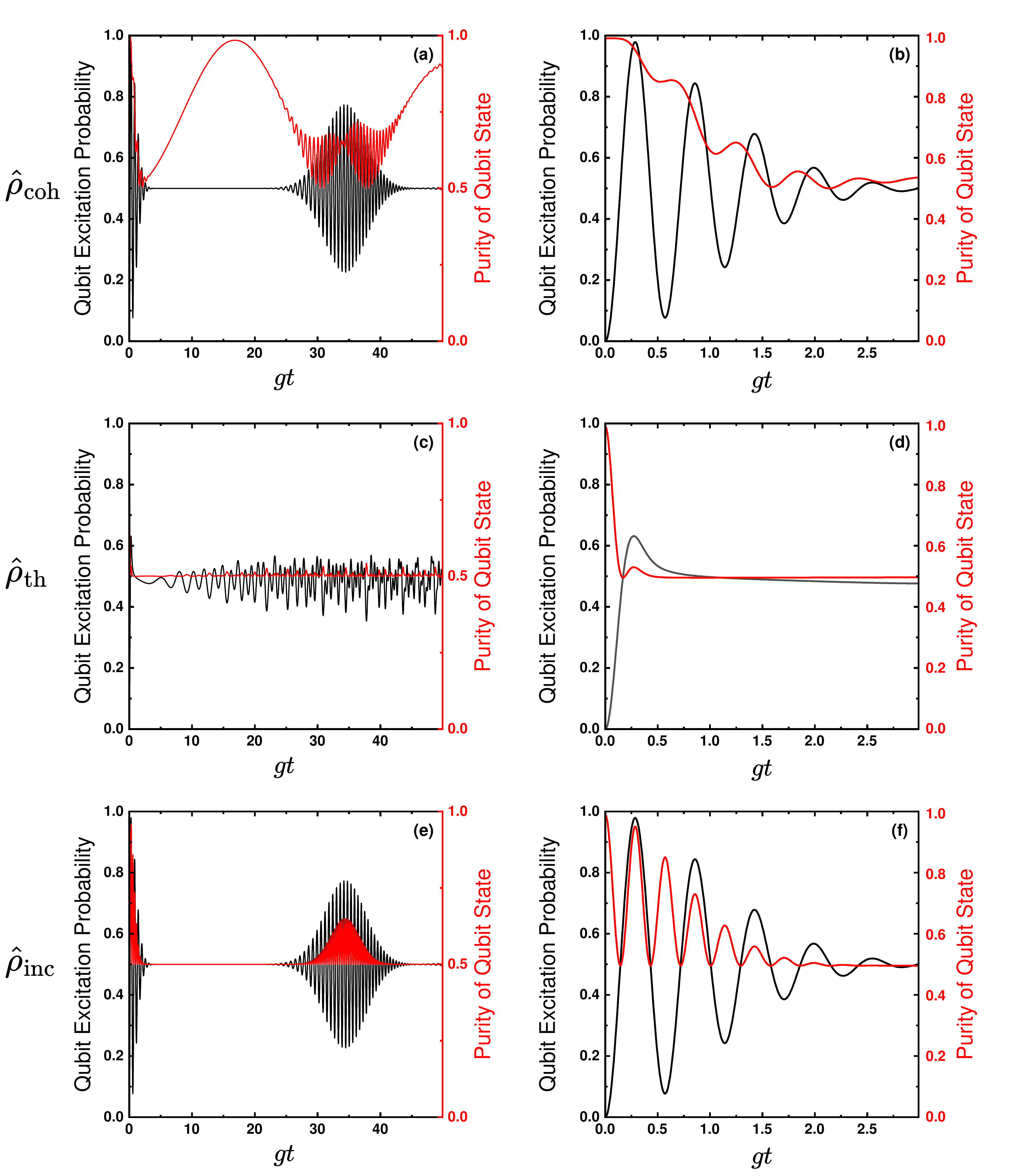}
        \caption{Time traces of the probability to find the qubit in the excited state (black line) and the purity of the atomic state (red line). Upper panels shows the results for a light field initially in a coherent state for $\langle N \rangle=30$. Middle panels show the results for a thermal state with $\langle N \rangle=30$. Bottom panels shows the results for a coherent state with $\langle N\rangle=30$ with all off-diagonal matrix elements removed. Panels on the right side show close-ups of the initial collapse of Rabi oscillations.}
        \label{fig:TimeTraces}
    \end{figure}

We will describe the light field and the qubit using density matrices $\hat{\rho}$ and $\hat{\rho}_\mathrm{q}$, respectively. 
To determine their joint dynamics, we solve the Liouville-von Neumann equation:
\begin{equation}
    \label{eq:Neumann}
\frac{d}{dt}\hat{\rho}_\mathrm{j}=-\frac{i}{\hbar}[\hat{H},\hat{\rho}_\mathrm{j}]
\end{equation}
for the joint density matrix $\hat{\rho}_\mathrm{j}=\hat{\rho}\otimes\hat{\rho}_\mathrm{q}$  for various initial states of the light field numerically. 
After obtaining the joint dynamics, we trace out the light field degrees of freedom to assess whether the qubit is in a pure or mixed state by calculating the purity of the reduced density matrix.
The time evolution of the probability of finding the qubit in its excited state, along with its purity, is illustrated in Figure \ref{fig:TimeTraces} for three different initial conditions of the light field.
At the top, we present the results for a light field with density matrix $\hat{\rho}_\mathrm{coh}$ that is initially in a coherent state containing 30 photons. 
This light field exhibits high coherence both from the perspective of resource theory and conventional optical second-order coherence.
We observe characteristic features of the non-trivial Jaynes-Cummings dynamics, including the collapse and revival of the qubit population, as well as the counterintuitive phenomenon whereby the qubit system approaches a pure state in the midst of the collapse region \cite{Eberly1980,Gea1990,Phoenix1991}.
At short times, we find typical damped Rabi oscillation behavior. 
Notably, at $gt\approx0.15$, the system reaches an almost pure superposition state, when the light field corresponds to a $\frac{\pi}{2}$-pulse.

We will now investigate how deviations from ideal quantum coherence manifest. 
We begin with the interferometric scenario, where quantum coherence is diminished solely due to additional mixedness arising from the presence of thermal light components. 
In the middle panel of Figure \ref{fig:TimeTraces}, we present the results for a light field with density matrix $\hat{\rho}_\mathrm{th}$ that is initially in a thermal state containing 30 photons and exhibits no quantum coherence.
This light field is incoherent both in terms of resource theory and conventional optical second-order coherence. 
The purity of the qubit state deteriorates rapidly, leading to an almost maximally mixed state. 
Notably, there are no revivals of the qubit occupation or purity within the collapse region. 
Although at $gt\approx0.15$ the probability of finding the qubit in its excited state reaches 0.5, the qubit remains in a maximally mixed state rather than a superposition state.
Consequently, $\hat{\rho}_\mathrm{th}$ proves to be a useless light field for coherent control.

It is illuminating to compare these two cases with the dephasing scenario, where quantum coherence is lost even when starting from an initially pure state. 
In the bottom panel of Figure \ref{fig:TimeTraces}, we present the results for a light field with density matrix $\hat{\rho}_\mathrm{inc}$, corresponding to a coherent state containing 30 photons, in which all off-diagonal elements of the density matrix are set to zero, resulting in $\mathcal C(\hat{\rho}_\mathrm{inc})=0$. 
This light field is fully incoherent from the perspective of resource theory but remains perfectly coherent in terms of conventional optical second-order coherence.
First, it is evident that the dynamics of the probability of finding the qubit in its excited state are identical to those observed for $\hat{\rho}_\mathrm{coh}$. 
We again observe typical collapse and revival regions for the qubit population. 
However, significant differences arise when examining the purity of the qubit state. 
At all times, the purity of the qubit does not exceed that observed for $\hat{\rho}_\mathrm{coh}$ and is lower in most cases. 
Furthermore, there is no revival of qubit purity during the collapse region.
Instead, the qubit remains in a maximally mixed state throughout this period.
Most importantly, at $gt\approx0.15$, while the probability of finding the qubit in its excited state reaches 0.5, the qubit state remains maximally mixed despite exhibiting identical population dynamics to those seen for $\hat{\rho}_\mathrm{coh}$. 
Consequently, $\hat{\rho}_\mathrm{inc}$ also proves to be useless as a light field for coherent control, even though it retains full coherence according to second-order optical coherence.

This is a remarkable finding because the process studied at $gt\approx 0.15$ corresponds to the excitation of a qubit in its ground state using a $\frac{\pi}{2}$-pulse, which is a standard procedure for preparing the qubit in a superposition state. 
However, we conclude that employing a classical $\frac{\pi}{2}$-pulse alone is insufficient for achieving this goal. 
Successful preparation of the qubit in a superposition state requires that the driving light field possesses quantum coherence.
Otherwise, one ends up with a mixed state that is not particularly useful for applications in quantum technologies.
This observation illustrates how our treatment of the resourcefulness of light fields generated by real lasers, as discussed in Section \ref{sec:LaserModel}, connects to relevant application scenarios. 
It is now instructive to conduct a quantitative analysis of the influence of quantum coherence on coherent control tasks. 

To this end, we will investigate the obtainable purity of the qubit when driven by a $\frac{\pi}{2}$-pulse for light fields with varying degrees of quantum coherence. 
We will consider both interferometric and dephasing scenarios, beginning with the former. 

\begin{figure}
        \centering
        \includegraphics[width=\textwidth]{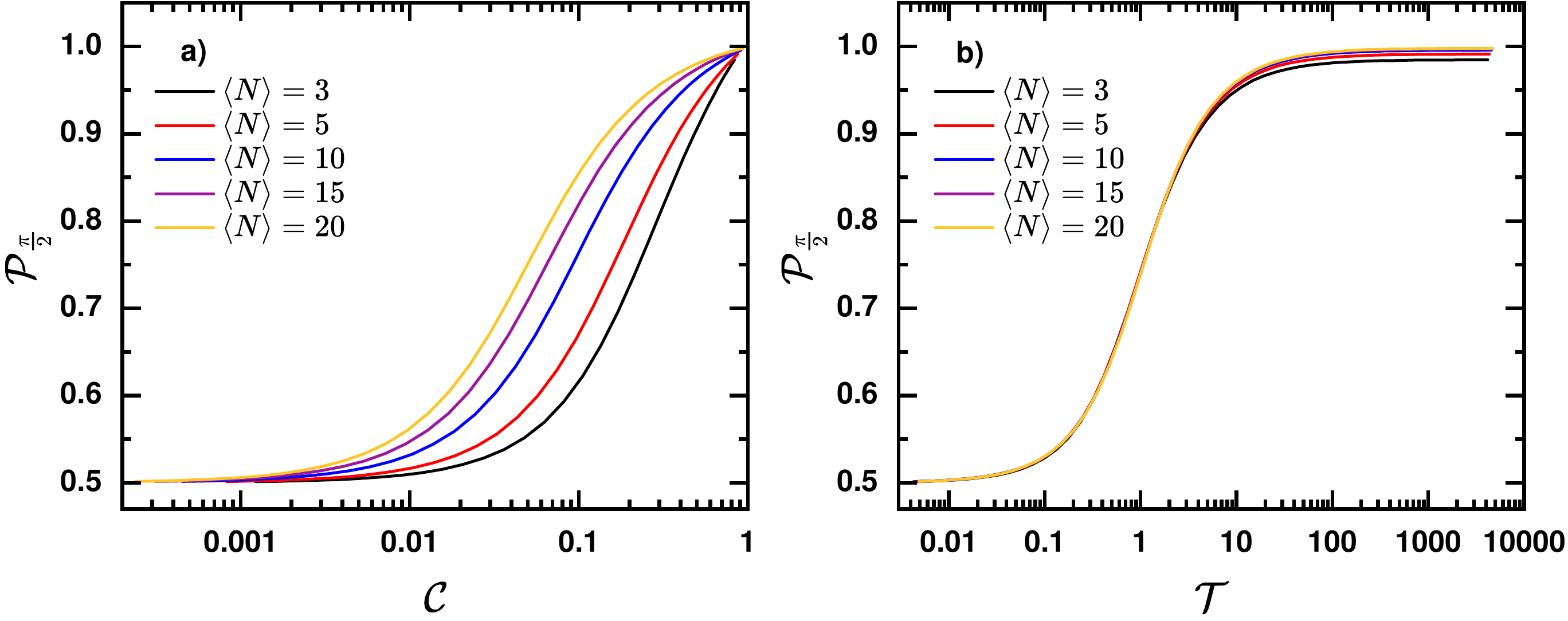}
        \caption{Qubit state purity following excitation with a $\frac{\pi}{2}$-pulse dependent on the states quantum coherence (a) and tightness (b) for light fields of varying total photon numbers of $\langle N \rangle$ of 3, 5, 10, 15 and 20.}
        \label{fig:PurityvsQCandTightness}
\end{figure}

Specifically, we will examine light fields with fixed total photon numbers $\langle N\rangle$, but varying fractions of thermal and coherent light. 
We will consider values of  $\langle N\rangle$ equal to 3, 5, 10, 15, and 20, determine the quantum coherence associated with these fields, and then calculate the purity of the qubit state following application of a $\frac{\pi}{2}$-pulse. 
The results are presented in Figure \ref{fig:PurityvsQCandTightness}(a).

\begin{figure}[h]
        \centering
        \includegraphics[width=\textwidth]{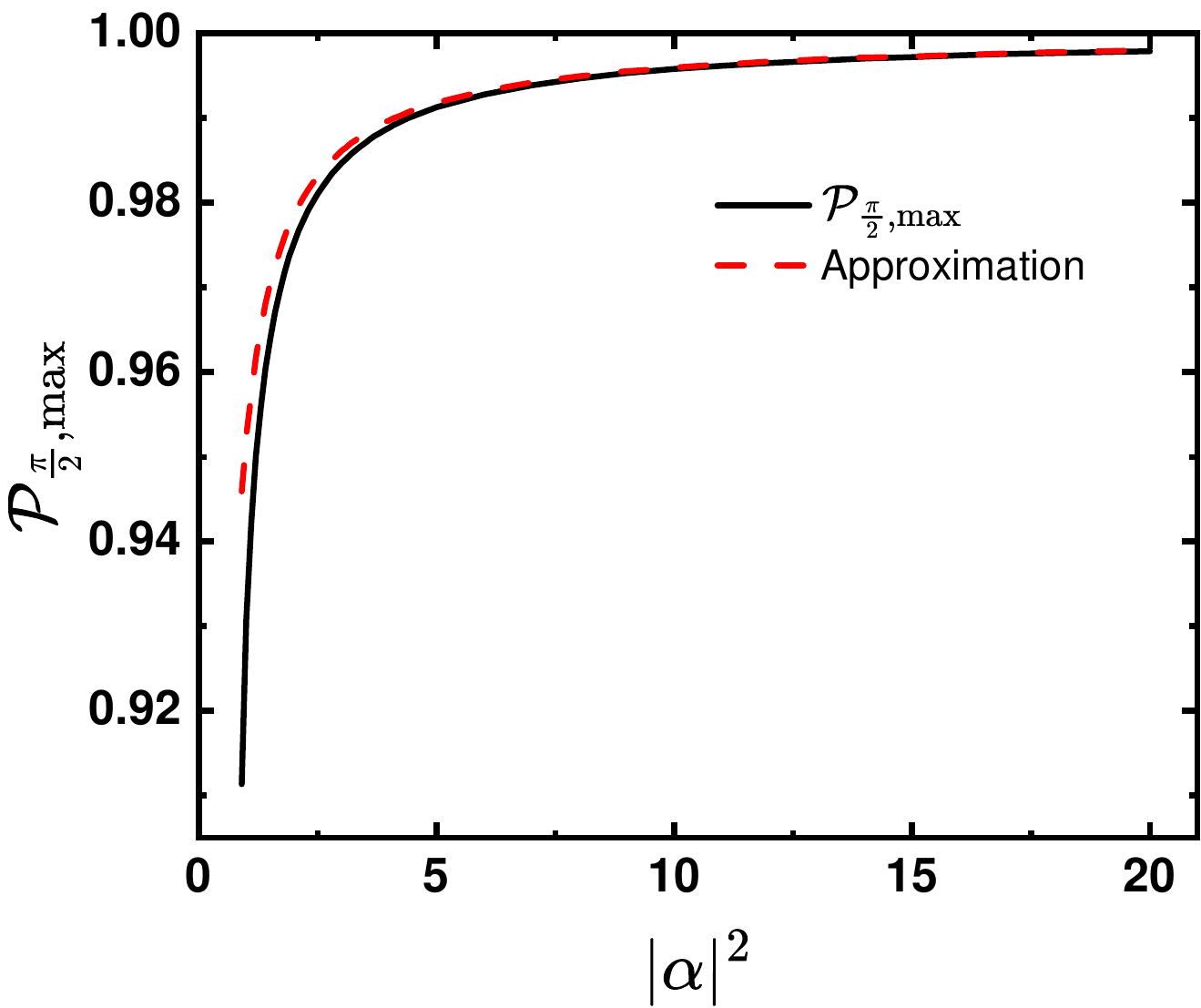}
        \caption{Maximal achievable qubit purity using a $\frac{\pi}{2}$-pulse of full coherent light of different mean photon number $|\alpha|^2$. The result of the simulation are given in black, the curve of a suitable approximation given in red. Further details are explained in the main text.}
        \label{fig:MaxPurity}
\end{figure}

For all photon numbers, we observe a significant increase in the achievable purity of the qubit superposition state with growing quantum coherence. 
Notably, this increase in purity begins at lower levels of quantum coherence for larger mean photon numbers. 
This can be intuitively understood by recognizing that the obtainable quantum coherence is constrained by the absolute thermal photon number.
Thus, increasing the mean photon number while keeping quantum coherence fixed effectively raises the coherent photon number.
As the mean photon number $\langle N\rangle$ increases, the pulse duration required to achieve a $\frac{\pi}{2}$ pulse becomes shorter, thereby reducing the overall impact of the noisy thermal component. 
This trend suggests that achieving coherent control becomes inherently more challenging at low photon numbers.
To provide a more comprehensive representation of the obtainable purity of the qubit superposition state, we will plot it against a more sophisticated quantity that incorporates the mean photon number. 
We will refer to this quantity as tightness $\mathcal{T}$, which depends on quantum coherence, mixedness, and $\langle N \rangle$. 
It is defined as follows:
\begin{equation}
\label{eq:Tightness}
 \mathcal T=\langle N \rangle \frac{ \mathcal C}{\mathcal M}.
\end{equation}
As the mixedness of a light field is solely determined by its absolute thermal photon number, increasing $\langle N\rangle$ while maintaining a constant coherent-to-thermal photon number ratio $\frac{|\alpha_0|^2}{\langle N\rangle_\mathrm{th}}$ does not lead to an indefinite increase in tightness $\mathcal{T}$, but rather results in a finite upper bound. A comprehensive examination of the mathematical properties related to tightness is beyond the scope of this study and will be addressed in future work.
Figure \ref{fig:PurityvsQCandTightness}(b) illustrates the obtainable purity of the qubit superposition state as a function of the driving light field's tightness for various values of $\langle N \rangle$.
Notably, the curves converge for low and medium tightness, following an almost universal trend.
However, for large tightness, the curves diverge slightly and approach a maximum that depends on $\langle N\rangle$, which is achieved for fully coherent excitation without any thermal component.
This observation leads directly to another key result of our study: the maximal achievable purity of the qubit superposition state depends not only on the purity of the driving light field, but also its photon number.
This relationship arises because $\mathcal P_\mathrm{inc}$ depends on the overlap between the light field state and the vacuum state, which diminishes significantly with increasing photon number.
The maximal obtainable purity across a broad range of photon numbers is depicted in black in Figure \ref{fig:MaxPurity}.
Below mean photon numbers of approximately $0.9$, the qubit does not reach an excited state probability of $0.5$, so we exclude mean photon numbers below this threshold from consideration.
For $0.9$ photons, we find an achievable qubit purity of around $0.912$. 
From there on the obtainable purity increases sharply with excitation power before it saturates for photon numbers above around 10.
For realistic mean photon numbers we find phenomenologically that the obtainable purity is limited by $\mathcal P_\text{inc}$ as follows:
\begin{equation}
    \label{eq:MaxPurity}
\mathcal P_{\frac{\pi}{2},\text{max}} = 1-0.5\, \mathcal P_\text{inc}^2.
\end{equation}

Only for very small mean photon numbers, we find a further degradation of the obtainable purity.
The additional reduction amounts to about $0.5\,\exp(-3 \cdot |\alpha|^2)$ and is caused by the fact that at such low photon numbers the standard deviation of the photon number exceeds the mean photon number itself, which renders the definition of a $\frac{\pi}{2}$-pulse complicated.
Still, for light fields applicable in realistic settings, the maximal obtainable qubit purity is described well by Eq.(\ref{eq:MaxPurity}).
These results highlight that the tightness defined in Equation (\ref{eq:MaxPurity}) serves as an excellent quantifier for the usefulness of a particular laser in coherent control tasks may also serve as a benchmark for optimization.

\begin{figure}[h]
        \centering
        \includegraphics[width=\textwidth]{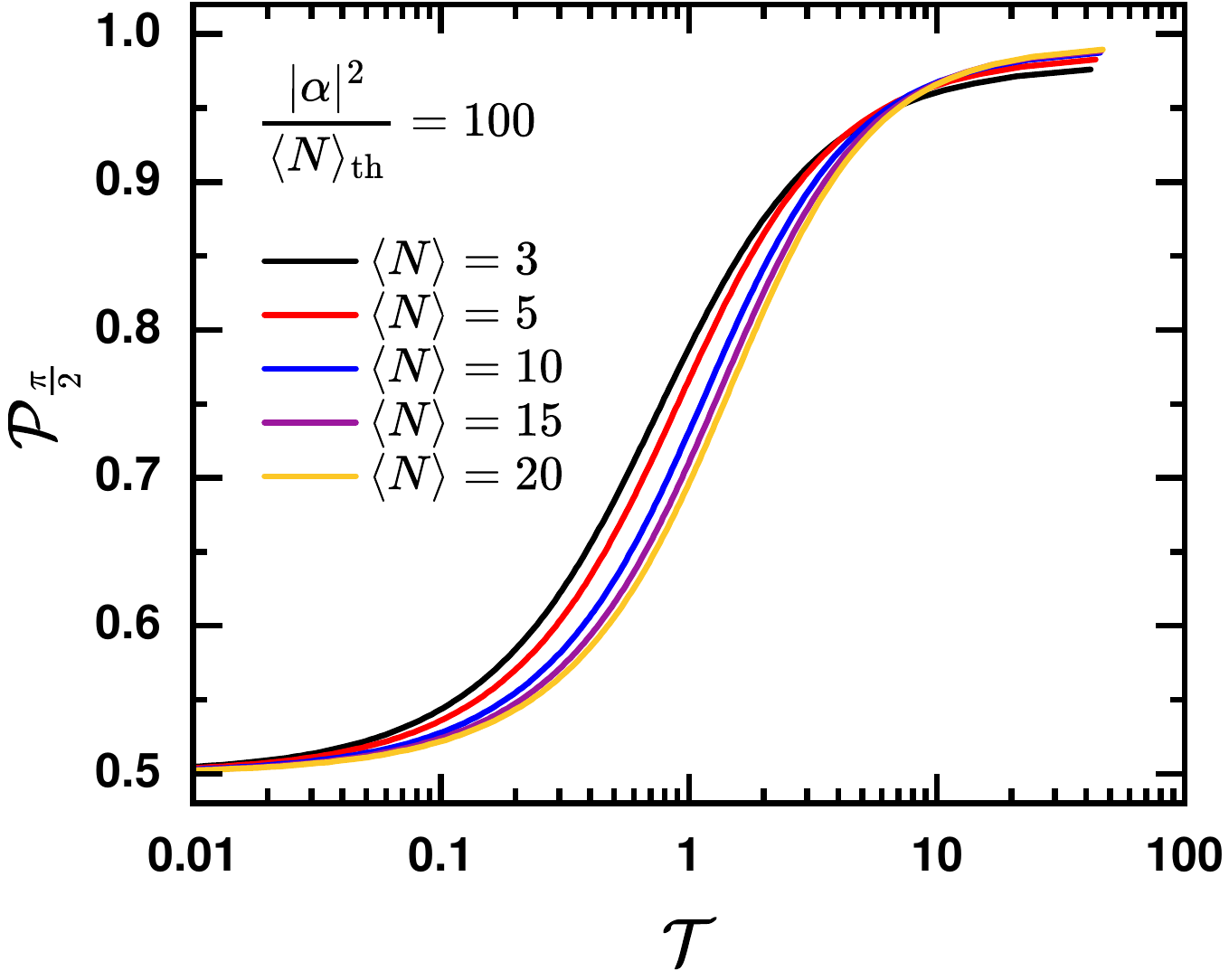}
        \caption{Qubit state purity following excitation with a $\frac{\pi}{2}$-pulse of different photon numbers but a common coherent to thermal ratio $\frac{\left|\alpha\right|^2}{\langle N \rangle_\text{th}} = 100$ and pure dephasing, dependent on the states tightness $ \mathcal T$.}
        \label{fig:DephasedState_N_Tightness}
\end{figure}

In the final part of our discussion, we investigate the dephasing scenario, in which quantum coherence may deteriorate due to interactions with the environment. 
Specifically, we do not only assess the maximum quantum coherence achievable by a state, but also explicitly incorporate the effects of dephasing.
To isolate the impact of the off-diagonal elements, we consider number-conserving scattering of a light field with an open bath resembling pure dephasing.
In this model, each photon lost to the bath is replaced by an equivalent photon from it \cite{Gardiner2004}.
Consequently, while the diagonal elements of the light field density matrix remain invariant, the off-diagonal elements decay over time. 
The time evolution of these density matrix element, denoted as $\hat{\rho}_\text{mn}$, is described by
\begin{align}
\hat{\rho}_\text{mn}\left( t \right) = \exp\left(-i \omega \left[m-n\right] t \right) \cdot \exp\left( -\gamma' \omega^2 \left[m-n\right]^2 t\right) \hat{\rho}_\text{mn}\left( 0 \right),
\end{align}
where $\hbar \omega$ corresponds to the photon energy and the effective damping constant $\gamma(\omega) = \gamma'\omega^2$ represents the scattering-rate with the bath.
The first term accounts for the conventional ultrafast oscillation of the coherences without altering their modulus, while the second term captures the decay of the modulus of the off-diagonal matrix elements.
Notably, this decay scales quadratically with the order of the off-diagonal.
This form of dephasing preserves the photon statistics and $\mathcal P_\text{inc}$; thus all lost quantum coherence is converted into additional mixedness.
To investigate how dephasing affects qubit initialization, we further analyze the purity $\mathcal P_{\frac{\pi}{2}}$ of the qubit's superposition state following a $\frac{\pi}{2}$-pulse.
In order to single out the influence of dephasing, we consider states with a fixed ratio of $\frac{|\alpha|^2}{\langle N \rangle_\text{th}}=100$ between the coherent and thermal contributions to the total photon number.
We consider different mean total photon numbers $\langle N\rangle$ and vary the magnitude of $\gamma t$.
We then determine how much the tightness $\mathcal{T}$ of the light field becomes reduced by dephasing.
The obtainable purity of the superposition state for a given dephasing-affected $\mathcal{T}$ is shown in Figure \ref{fig:DephasedState_N_Tightness}.
For comparison, we also show the obtainable purity with respect to $\gamma t$ in Appendix \ref{app:dephasing}.
As noted before, the maximal achievable $\mathcal{P}_{\frac{\pi}{2}}$ in the absence of any dephasing increases with larger $\langle N \rangle$ due to reduced $\mathcal{P}_\text{inc}$.
With increasing dephasing $\mathcal{T}$ decreases and $\mathcal{P}_{\frac{\pi}{2}}$ drops along with it.
The curves do not show universal behavior in the presence of dephasing, but in all cases the middle of the threshold region arises for $\mathcal{T}$ close to unity.
The curves cross, so that lower $\langle N \rangle$ actually results in larger obtainable $\mathcal{P}_{\frac{\pi}{2}}$ for partially dephased light fields of the same $\mathcal{T}$.
It may seem surprising at first that light fields with lower photon number may show a better performance for this task, but it has been shown before that splitting a light field may actually increase its quantum coherence \cite{Ares2023, Diez2024}.
For the task at hand, the increased performance can be explained qualitatively by two effects.
First, reducing $\langle N \rangle$ also directly reduces the thermal photon number proportionally, which has a strong influence on the mixedness of the light field and second, we consider a two-level system that can only absorb one photon, so only the first-order off-diagonal elements influence the obtainable purity.
Interestingly, this behavior provides another way to optimize qubit initialization and operation by adjusting the mean photon number of the light fields if the dephasing has been characterized properly for a given experimental setting. 
Supplementary to the previous results we also investigate the obtainable purities dependence on the first off diagonals only in Appendix \ref{app:C1}.

\section{Conclusions and Outlook}
In summary, we have developed a conceptually straightforward resource-theoretical model for the formation of quantum coherence in non-ideal realistic lasers and similar light fields relevant to applications in photonics. 
We demonstrated that achievable quantum coherence is constrained by both the absolute residual thermal photon number and the purity of the dephased light field. 
Furthermore, we illustrated how the quantum coherence of a light field directly influences the obtainable purity of a superposition state when initializing a qubit, while conventional coherence quantifiers such as $g^{(2)}(0)$ fail to predict any differences.
Most importantly, we introduced easily implementable measurement schemes for determining the quantum coherence of displaced thermal states without requiring a phase reference.
Therefore, we established a powerful benchmark for optimization of laser sources and integrated quantum technologies.

While our approach has focused on a single basic task to illustrate the effectiveness of our framework, extending this study to examine the influence of quantum coherence on more sophisticated excitation and population transfer schemes as well as other relevant protocols in photonics \cite{Vitanov2017, Jaksch2000, Schwartz2016, Bracht2021} will be an intriguing future endeavor. 
Additionally, revisiting pioneering work on qubit control with coherent fields \cite{Gea2002, Igeta2013} while explicitly considering realistic laser sources that exhibit deviations from perfect coherence will also be worthwhile.
Generally, it seems beneficial to incorporate resource theories at the design stage of light sources for photonic quantum technologies to adequately account for their imperfections. 
With our work, we have laid the groundwork for establishing resource theories that are both easy to understand and implement experimentally within communities working on laser and general light source design and manufacturing.

We hope that our study serves as a starting point for extending resource theoretical studies of concrete technological tasks such as interferometry \cite{Biswas2017, Masini2021} particularly regarding characteristic imperfections associated with real world light sources.
To this end, it will also be very interesting to investigate the quantum coherence properties of non-conventional coherent light sources such as Bose-Einstein condensates of photons \cite{Klaers2010} or polaritons \cite{Kasprzak2006} or free-electron lasers \cite{Pellegrini2016}.
Finally, it also seems promising to consider spectroscopic scenarios, where the degree of quantum coherence of an emitted light field reveals nontrivial properties of the emitting system.
Inherently quantum features have been identified in seemingly classical processes such as Cherenkov radiation before by properly studying the interplay between the quantum coherence of the emitter and the conventional coherence of the light field \cite{Karnieli2021}, and we envision that studying the quantum coherence of the emitted light field can provide further fascinating insights. 

\appendix
\section{\label{app:dephasing} Obtainable purity with respect to the dephasing constant}
\begin{figure}[h]
        \centering
        \includegraphics[width=\textwidth]{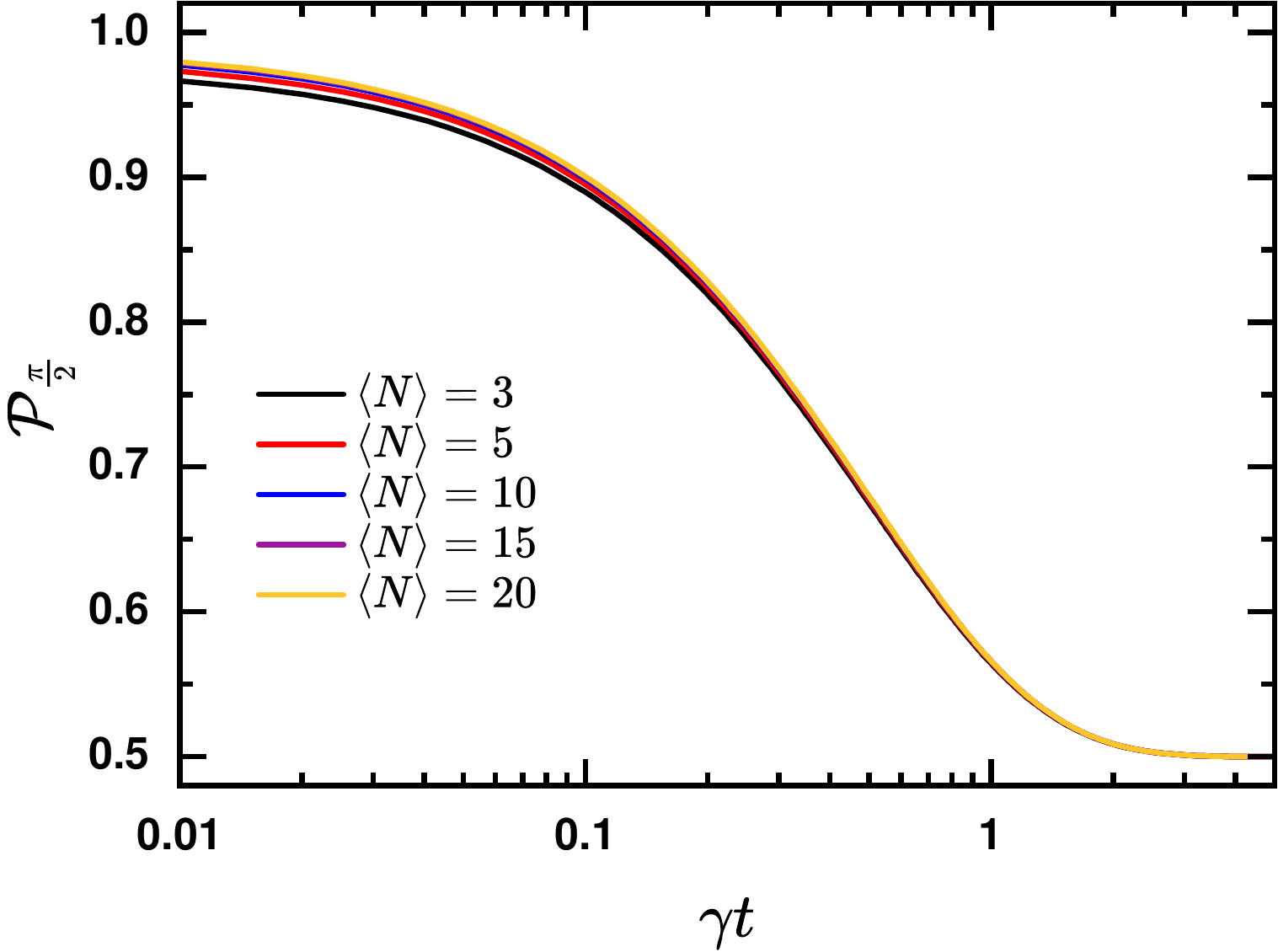}
        \caption{Qubit state purity for various levels of dephasing $\gamma t$ following excitation with a $\frac{\pi}{2}$-pulse of different photon numbers but a common coherent to thermal ratio $R = \frac{\left|\alpha\right|^2}{\langle N \rangle_\text{th}} = 100$.}
        \label{fig:DephasedState_N_Gammat}
\end{figure}
We complement our analysis of the dephasing scenario by investigating the obtainable purity with respect to the dephasing constant $\gamma t$. 
To ensure comparability with our previous results, we utilize the same states as described in the main text.
The corresponding curves are illustrated in Fig. \ref{fig:DephasedState_N_Gammat}. 
In the absence of dephasing, all curves are distinct, with achievable purity increasing with the mean photon number. 
However, as we introduce dephasing, the obtainable qubit purity steadily declines, resulting in converging curves. 
By a dephasing time of $\gamma t = 1$, the curves have merged, and the obtainable purity has decreased to approximately $0.55$ only.

\section{\label{app:C1} Coherence reduction via thermal noise and dephasing with respect to the first order off-diagonal elements}
\begin{figure}[h]
        \centering
        \includegraphics[width=\textwidth]{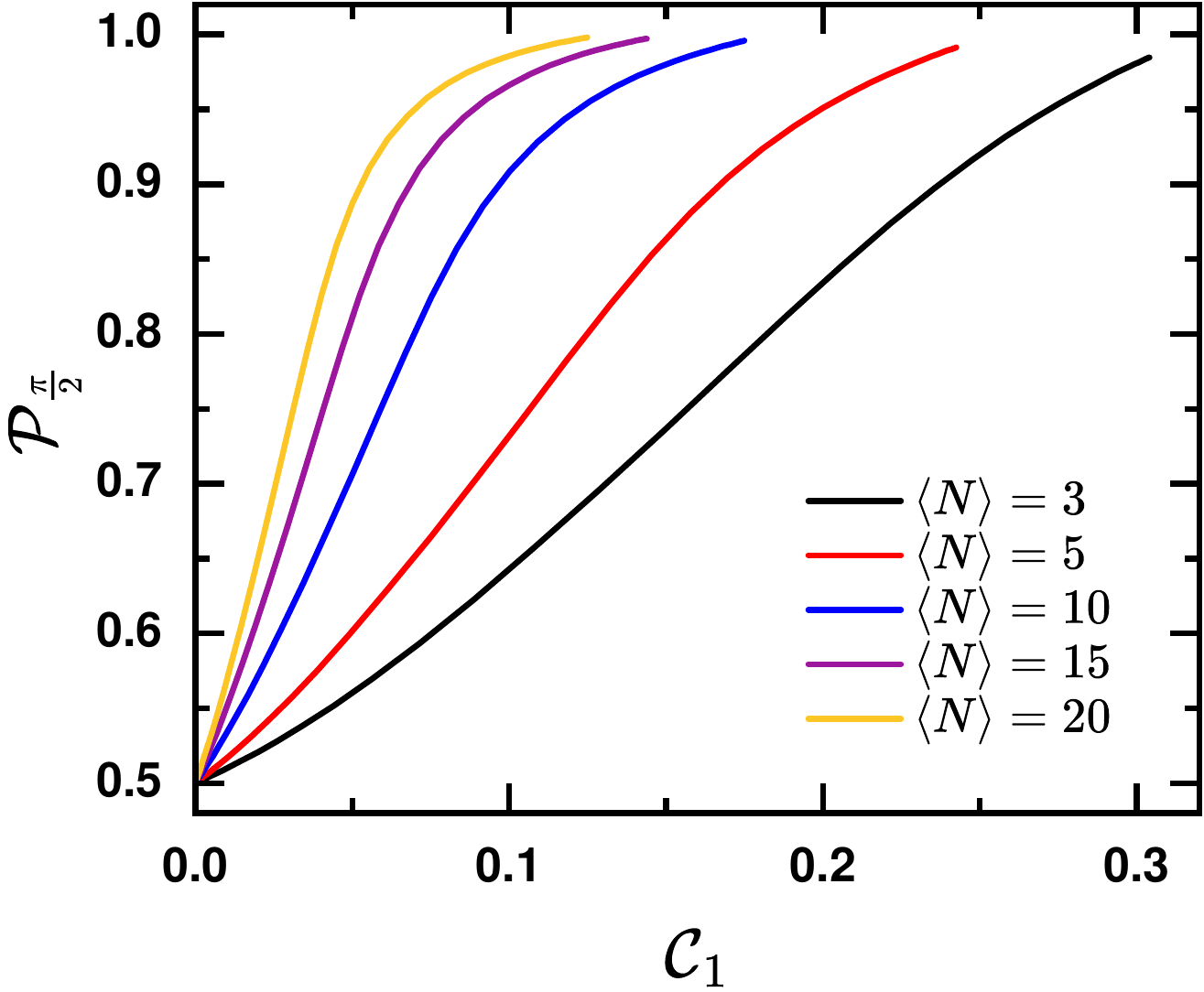}
        \caption{Qubit state purity following excitation with a $\frac{\pi}{2}$-pulse of different photon numbers and a varying coherent to thermal ratio $\frac{\left|\alpha\right|^2}{\langle N \rangle_\text{th}}$, dependent on the quantum coherence of the first order off-diagonals $\mathcal C_1$.}
        \label{fig:Ratio_N_FirstOffQC}
\end{figure}

\begin{figure}[h]
        \centering
        \includegraphics[width=\textwidth]{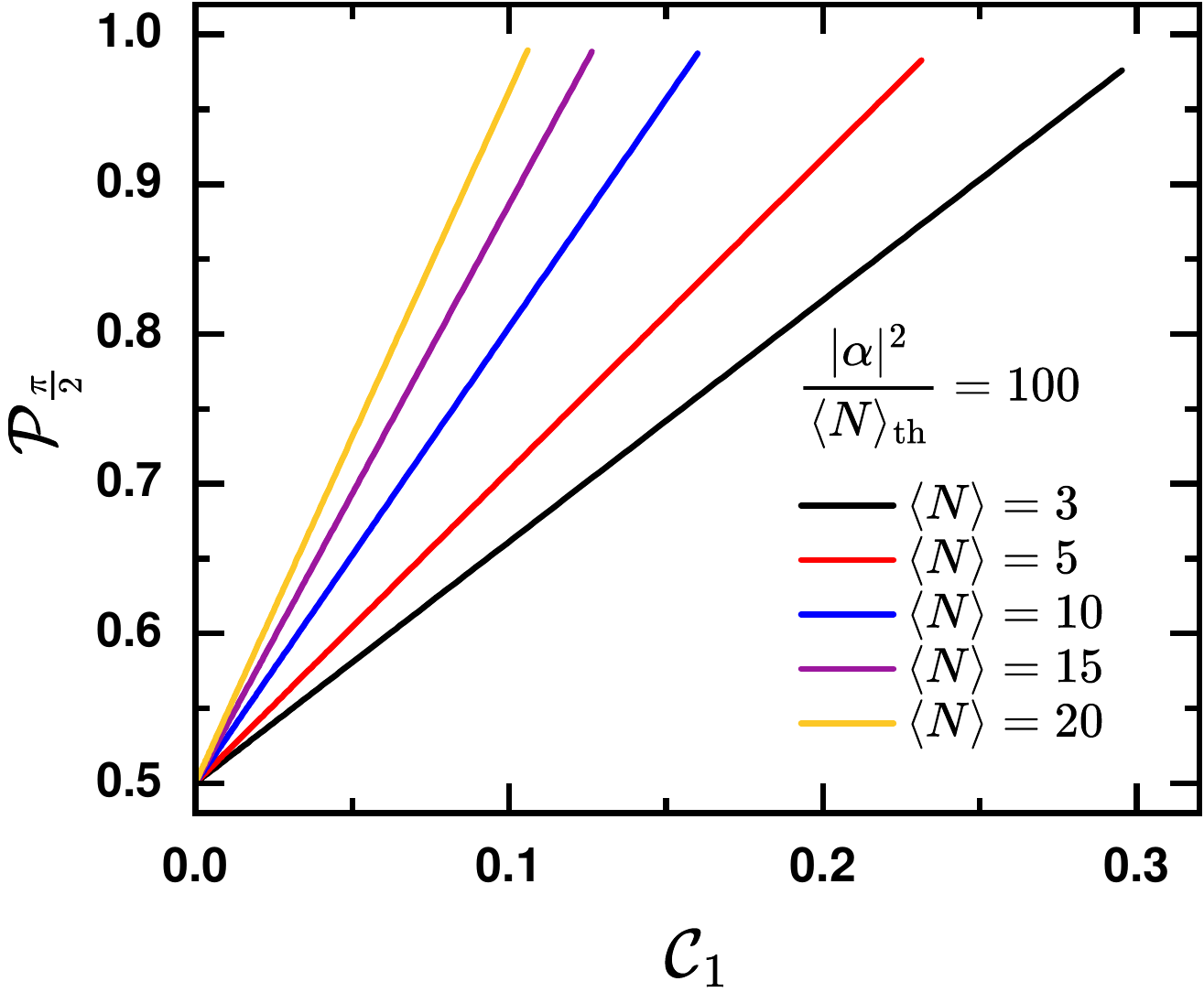}
        \caption{Qubit state purity following excitation with a $\frac{\pi}{2}$-pulse of different photon numbers but a common coherent to thermal ratio $\frac{\left|\alpha\right|^2}{\langle N \rangle_\text{th}} = 100$ and pure dephasing, dependent on the quantum coherence of the first order off-diagonals $\mathcal C_1$.}
        \label{fig:Dephasing_N_FirstOffQC}
\end{figure}
Since we are considering the coupling of a light field to a two-level system, only the main diagonal and the first off-diagonal elements of the light field's density matrix significantly influence the qubit dynamics. 
The probability of qubit excitation depends on the photon statistics, while the purity of the state is also affected by the first-order off-diagonal elements. 
In addition to general quantum coherence $\mathcal C$, the quantum coherence $\mathcal C_1$ that arises solely from the first order off-diagonals of the density matrix is therefore also a quantity of interest:
\begin{align}
\mathcal C_1 = \sum_{m \in \mathbb{N}} \left[\lvert\hat{\rho}_{m,m - 1}\rvert^2 +\lvert\hat{\rho}_{m,m + 1}\rvert^2\right]\text{\,.}
\end{align}
We will now reconsider the coherence losses due to thermal components and dephasing with respect to $\mathcal C_1$. 
For the thermal components, we consider light fields of fixed mean photon number and vary the ratio between the coherent and thermal contributions to the photon number.
As the thermal contribution increases, the magnitude of the first-order off-diagonal matrix elements becomes reduced.
We then plot the obtainable purity of the qubit superposition state against the remaining first-order quantum coherence.
For dephasing, we directly consider the dephased density matrices and also plot the obtainable purity of the qubit superposition state against the remaining quantum coherence of the first-order off-diagonal matrix elements.
The corresponding curves for both scenarios are illustrated in Figures \ref{fig:Ratio_N_FirstOffQC} and \ref{fig:Dephasing_N_FirstOffQC}. 
In the case of thermal noise, the curves for all photon numbers exhibit a nonlinear increase with first-order quantum coherence, but eventually approach maximum purity as thermal noise diminishes. 
Both the gradient of this initial increase and the width of the saturation plateau expand with increasing photon number. 
Notably, the required $\mathcal C_1$ to obtain close to ideal qubit state purity decreases for larger photon numbers. 
This relationship contrasts with that observed for $\mathcal C$ itself and can be traced back to the fact that off-diagonal matrix elements beyond the first order become more prominent for larger mean photon numbers.

At first glance, the curves considering the effect of dephasing appear similar compared to those from the previous scenario.
However, the relationship between obtainable purity and $\mathcal C_1$ becomes linear in this case. 
Ultimately, this allows us to derive the purity achieved in the dephasing scenario from results obtained in the interferometric scenario.

\begin{acknowledgments}
We gratefully acknowledge support through the QuantERA II Programme that has received funding from the EU H2020 research and innovation programme under GA No. 101017733, and by the Deutsche Forschungsgemeinschaft (DFG) within the projects under GA No. 532767301.
\end{acknowledgments}

\section*{Data Availability}
Data that support the findings of this article are openly available \cite{DataRepository}.

\bibliography{bibliography}

\end{document}